# Material selection for mid-infrared thin-film coatings and windows

Jin-Woo Cho, Tanuj Kumar, Hongyan Mei, and Mikhail A. Kats*

*Department of Electrical and Computer Engineering, University of Wisconsin-Madison, Madison, Wisconsin 53706, USA*
**Corresponding author: mkats@wisc.edu*

## Abstract

We summarized the room-temperature optical properties for infrared-transparent materials, defining transparency windows for two different applications: thin-film coatings (absorption coefficient $\alpha < 10$ cm$^{-1}$) and windows ($\alpha < 1$ cm$^{-1}$). The transparency requirements for thin films are substantially less stringent, enabling the use of many more optical materials for a given wavelength range. To make an easy-to-use reference, we categorized materials by chemical group, crystallinity, and typical deposition technique, and discussed practical pros and cons such as chemical and thermal stability, mechanical hardness, and cost. The datasets and plotting scripts are provided so that users can regenerate transparency-window maps for different applications. If you do optical design at infrared wavelengths, we recommend that you print out the figures and stick them on your wall.

## Introduction

Optical components in the mid-infrared (mid-IR) spectral range, typically wavelengths ($\lambda$) from 2 to 25 μm, are widely used in thermal imaging [1], chemical and biomedical sensing [2], free-space communications [3], and laser-based manufacturing [4]. Optical components such as windows, filters, polarizers, and lenses, as well as the thin-film coatings for all of these components, typically need to be as transparent as possible, and therefore should be made from materials that have the lowest optical loss.

No mid-IR material is perfect, and tradeoffs must be made in material selection for each type of optical component in various applications. To aid material selection, many references compile lists of IR-transparent materials [5], [6], [7], [8], [9], [10], mostly considering transparency through IR *windows*, which have thickness on the order of a millimeter. However, transparency is not a binary characteristic, and in many situations, materials can be used even if they are too lossy for millimeter-thick windows.

For example, a comprehensive reference of optical materials from 30 years ago lists the transparency edge of silicon (Si) as 6.5 μm [8]; meanwhile, Si wafers with thickness < 500 μm are widely used today as transparent substrates for optical metasurfaces [11], [12], and other devices in the mid-IR (8–14 μm) [13]. Many optical devices today use even thinner substrates, for example Si or silicon nitride ($Si_3N_4$) membranes which can be submicron in thickness [14], [15]. Therefore, the definition of "transparent" or "low loss" depends strongly on the application domain, for example the thickness of a film or window or the quality factor of resonators.

We carefully reviewed the literature for measured optical properties (i.e., $n$ and $\kappa$) of materials available for mid-IR applications, and summarized the data in an easy-to-use format, separating the use cases of thin-film optical coatings (**Fig. 1**) and macro-thickness optical windows (**Fig. 2**). These two use cases differ in the acceptable levels of optical loss, as well as in the available growth techniques. For example, single-crystal Si windows are typically made using Czochralski growth [16], whereas thin films are deposited using physical or chemical vapor deposition, and can be made amorphous (a)- or polycrystalline (poly)-, with single-crystal (c)-Si thin films only available via special processes like epitaxial liftoff [17].

To enable consistent comparison across all materials, we defined quantitative internal-transparency criteria for both thin-film coatings and thick windows. For thin films, we set as the threshold an absorption coefficient of $\alpha < 10$ cm$^{-1}$, corresponding to > 95% internal transmittance at 10 μm thickness (**Fig. 1**).

Although 10 μm is considerably thicker than individual layers of typical thin-film stacks, this threshold provides a sufficient margin to ensure negligible absorption in many thin-film coatings. For thick optical windows and refractive optical components, we applied a stricter criterion of $\alpha < 1$ $cm^{-1}$, corresponding to internal transmittance of > 99% at 500 μm thickness (**Fig. 2**). 500 μm is a common wafer thickness, and is thinner than a typical window. We note that the $\alpha < 1$ $cm^{-1}$ criterion was also used by W. J. Tropf, M. E. Thomas, and P. Klocek in an excellent survey of IR materials from 1992 [18], and we view the present review paper as a focused update of that paper.

Using these criteria for optical absorption, we grouped the materials by their chemical composition, crystallinity, and deposition technique, and plotted their refractive-index dispersion over their regions of transparency across the 0.2–25 μm spectral range, in **Fig. 1** for thin-film coatings and **Fig. 2** for optical windows. In an appended Zenodo record [19], we are including all of the datasets and a tool that allows the user to dynamically generate similar figures with an arbitrary α threshold.

**Infrared transparent materials for thin-film coatings**

While in principle any material could be used as a thin-film coating, in practice some materials are much easier to use than others given available deposition techniques, material compatibility, etc. For example, it is straightforward to deposit films of amorphous germanium (a-Ge) using physical vapor deposition [20], but using single-crystal Ge (c-Ge) as a layer in a thin-film stack requires a sophisticated epitaxial liftoff and transfer process [21], [22].

In **Fig. 1**, we summarize materials that can readily be used in thin-film coatings, with typical deposition techniques labeled from [a] to [g] in order of increasing epitaxial quality, ranging from evaporation [a], which is mostly substrate-agnostic, to molecular-beam epitaxy (MBE) [g], which usually needs lattice-matched substrates to grow. These materials are categorized into six representative chemical groups:

**(i) Fluoride insulators (poly-$MgF_2$, poly-$CaF_2$, poly-$BaF_2$)**

Fluoride materials are transparent from the ultraviolet (UV) range to different cutoffs in the mid-IR, in particular to $\lambda < 8.7$ μm for magnesium fluoride (poly-$MgF_2$) [23], $\lambda < 12$ μm for calcium fluoride (poly-$CaF_2$) [24], and $\lambda < 16$ μm for barium fluoride (poly-$BaF_2$) [25], with the progressively longer cutoffs reflecting the increasing cation atomic mass that shifts phonon modes further to longer wavelengths. In addition, they possess low $n \sim 1.3$–$1.5$ due to their large band gaps, which arise from the strongly ionic bonds driven by fluorine's high electronegativity [26].

These materials are typically polycrystalline when deposited by evaporation or sputtering [27], [28], making them well-suited for multilayer coatings across a broad range of wavelengths. However, their mechanical softness and sensitivity to moisture can be drawbacks [18], often requiring the use of additional protective layers.

**(ii) Oxide insulators (a-$SiO_2$, a-$Al_2O_3$, a-$HfO_2$, a-$TiO_2$)**

Oxide materials are the most widely used transparent insulators, with $n$ ranging from 1.4 to 2.4. Their transparency window is limited to $\lambda < 5$ μm for silicon dioxide (a-$SiO_2$) [29] and titanium dioxide (a-$TiO_2$) [29], $\lambda < 8$ μm for aluminum oxide (a-$Al_2O_3$) [30], and $\lambda < 10.5$ μm for hafnium dioxide (a-$HfO_2$) [31], due to multi-phonon absorption [32]. Note that the 2–5 μm range is not normally considered to be a transparent range for a-$SiO_2$, but the losses are sufficiently small for thin-film coatings, given our criteria. Note also that the transparency window of a-$SiO_2$ is interrupted at $\lambda \sim 2.8$ μm by −OH absorption [33], which typically arises from residual hydroxyl groups incorporated during low-temperature deposition. High temperature annealing (> 900 °C) is a common approach to reduce this absorption [34]; however, it may induce defect formation (e.g., voids) [35] and interfacial delamination.

Most oxides can be deposited in amorphous form by evaporation, sputtering, atomic layer deposition (ALD), and/or chemical vapor deposition (CVD). This deposition flexibility combined with their excellent chemical stability and mechanical hardness leads them to be extensively used for visible and near-IR (~0.8 to 2 μm) optical applications, and the losses remain small enough to extend into the mid-IR. While multi-phonon absorption limits their use at longer mid-IR wavelengths, the same phonon absorption can be exploited in some applications, for example to enhance emissivity for infrared emitters [36].

**(iii) Nitride insulators (a-SiN, a-AlN)**

Amorphous silicon nitride (a-SiN) [37], [38] exhibits a moderately high $n \sim 2.0$ with a transparency window up to $\lambda \sim 7$ μm (**Fig. 1**), and its refractive index and transparency window can be substantially tuned through composition control (e.g., Si-rich or N-rich films) [39].

Plasma-enhanced CVD (PECVD, ~300–400 °C) [39] and reactive sputtering [40] are commonly used to make SiN optical coatings, both generally forming non-stoichiometric $SiN_X$ films that are amorphous. PECVD SiN films exhibit an absorption feature near $\lambda \sim 2.9$ μm due to the –NH phonon mode of residual hydrogen from the deposition process [41], which interrupts the transparency window. Reactive sputtering typically incorporates less hydrogen than PECVD because it requires no hydrogen-containing precursors, resulting in reduced optical loss [40]. Note that the stoichiometry of $SiN_X$ films is typically determined by both gas composition and deposition temperature, with lower temperatures leading to incomplete reactions and hydrogen incorporation. At higher deposition temperatures, low-pressure CVD (LPCVD, ~850 °C) can yield near-stoichiometric poly-$Si_3N_4$ with improved optical uniformity. Post-deposition annealing (~1300 °C) can further approach fully stoichiometric poly-$Si_3N_4$ [37], enabling ultra-low optical loss [42]. However, such high temperatures induce thermal stress and interfacial diffusion in thin film platforms.

Aluminum nitride (a-AlN) [30], [43], that can also be deposited via PECVD and reactive sputtering, exhibits a high $n \sim 2.1$ with a transparency window from 1.2 to 3.3 μm but offers superior thermal/chemical robustness than SiN [44]. Although it requires higher post annealing temperatures (> 1500 °C) for optimal crystallinity [45], its strong Al–N bond and the absence of competing stable aluminum nitride phases yields intrinsically stable stoichiometry [44], making deposition reproducible and reliable.

**(iv) Chalcogenide semiconductors (poly-ZnS, poly-ZnSe, c-CdTe)**

Chalcogenides combine chalcogen group elements (group 16 in the periodic table), such as sulfur (S), selenium (Se), and tellurium (Te), with chalcophile elements such as zinc (Zn), arsenic (As), cadmium (Cd), or antimony (Sb). These materials show moderately high $n \sim 2.2$–$2.7$ and exhibit broad IR transparency due to their heavy constituent atoms, which shift the multi-phonon absorption edge to longer wavelengths. Therefore, the mid-IR transparency cutoff extends from $\lambda \sim 17$ μm for poly-ZnS [46], [47] to $\lambda < 22$ μm for poly-ZnSe [25], to $\lambda > 25$ μm for c-CdTe [48], becoming longer with increasing constituent atomic masses.

In particular, ZnS and ZnSe are widely used as IR-transparent coating materials. ZnS has a transparency window of $\lambda = 0.4$–$17$ μm, whereas ZnSe provides a broader transparency range of $\lambda = 0.5$–$22$ μm and a slightly higher $n$ (~ 2.2 for ZnS and ~2.4 for ZnSe). Their intermediate refractive indices are well-suited for antireflection coatings between high-$n$ substrates (e.g., Si, Ge, or GaAs) and air. Both ZnS and ZnSe can be readily deposited by evaporation and sputtering, typically forming polycrystalline films. ZnS offers high mechanical hardness and environmental durability, while ZnSe is mechanically softer and typically requires a protective coating when used in harsh environments [49].

While CdTe is a promising candidate for mid-IR coatings ($n$ ~2.0 to 3.0) due to its broad transmission up to 25 μm, it is less-commonly used due to its the toxicity of Cd, and the brittleness of the compound, which limits the area over which CdTe can be reliably deposited. Consequently, efforts to replace Cd with less toxic alternatives are ongoing; for example, tin (Sn) has been explored as a replacement, leading to SnTe coatings [50], [51].

**(v) Group III-V semiconductors (c-GaP, c-GaAs)**

III-V materials, like gallium phosphide (c-GaP) and gallium arsenide (c-GaAs), provide high $n \sim 3.0–3.5$ and relatively wide IR transparency (given the high index) of $\lambda < 17$ μm for c-GaP [52] and $\lambda < 18$ μm for c-GaAs [53]. These materials can be deposited by molecular beam epitaxy (MBE) or metal-organic CVD (MOCVD) on lattice-matched substrates, yielding single-crystalline films. However, the lattice-matching requirement together with high growth temperatures means that these materials can typically only be incorporated onto other III-V semiconductor stacks, particularly in light-emitting devices [54] and solar cells [55].

Transfer processes for epitaxially grown c-GaP and c-GaAs may enable thin-film optics, but they still face limitations in scalability for large-area manufacturing [56], [57]. Polycrystalline GaP and GaAs films can also be deposited on lattice-mismatched substrates such as glass by RF sputtering [58], [59], but non-stoichiometry and high defect densities yield substantial optical losses [58], limiting their use as IR-transparent coatings.

**(vi) Group IV semiconductors (c-Si, a-Si, c-Ge, a-Ge)**

Group IV semiconductors exhibit very high $n \sim 3.4–4.0$, as shown in **Fig. 1**. The c-Si data in **Fig. 1** corresponds to a low doping level [60], exhibiting transparency from 1 μm to beyond 25 μm. Note that although c-Si has a primary Si–Si phonon mode near $\lambda \sim 16$ μm [61], this mode is mostly inactive due to the inversion symmetry of the diamond cubic lattice [62], keeping the resulting absorption below the thin-film threshold defined in this work. In contrast, a-Si shows a shorter IR transparency window below $\lambda \sim 10$ μm [63]. This reduction is primarily due to structural disorder, which breaks the inversion symmetry and broadens and enhances the Si–Si phonon absorption, as well as Si–H vibration modes arising from hydrogen incorporated during deposition [64]. Additionally, Si-H absorption also contributes to a gap in the transparency window of a-Si near $\lambda \sim 5$ μm [64].

a-Ge [65] is one of the most-common IR-transparent materials, combining a broad transparency window of $\lambda = 2–25$ μm, low dispersion, and mechanically and chemically robustness. Unlike a-Si, a-Ge shows comparable transmittance window with c-Ge [66] because its heavier atomic mass shifts phonon vibration modes toward much longer wavelengths [61]. Consequently, even with disorder-induced activation and broadening of phonon modes, the primary mid-IR transparency window of Ge remains unaffected.

While a-Si and a-Ge films can be deposited by evaporation, sputtering or CVD, c-Si and c-Ge generally require high-temperature CVD [16] or epitaxial growth [22], which limits substrate compatibility, though epitaxial transfer techniques can be used when needed. In particular, c-Si has lower loss and therefore a broader transparency window compared to a-Si, and there exist several methods for the transfer of c-Si films from Si-on-insulator (SOI) wafers onto arbitrary substrates [67].

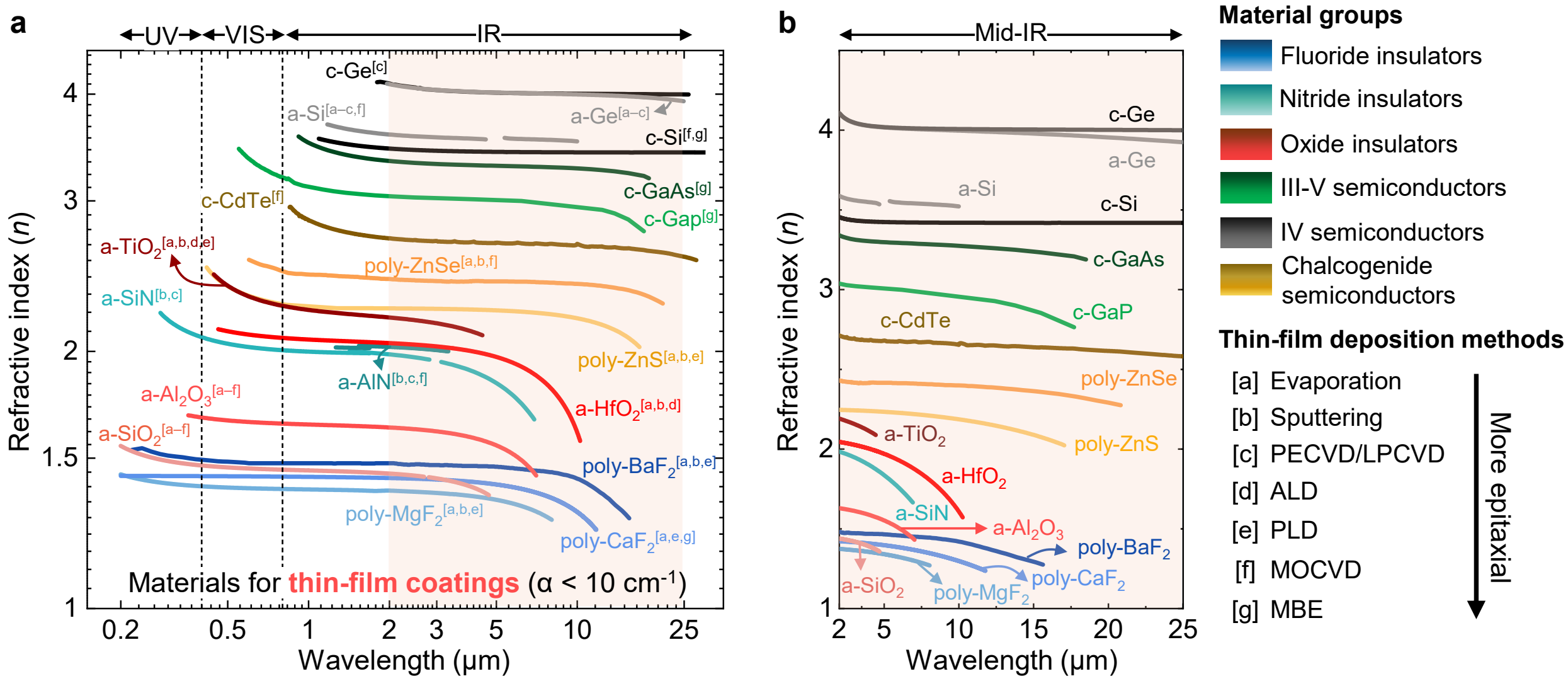

**Fig. 1. Infrared-transparent materials for thin-film coatings.** (**a**) Real part of the refractive index ($n$) for various candidate materials, where the wavelength regions satisfying an absorption coefficient of $\alpha$ < 10 $cm^{-1}$ are defined as their 'transparency window' and plotted over the 0.2–25 μm spectral range. The materials are categorized by their chemical group. Typical thin-film deposition techniques are indicated for each material, arranged from less-epitaxial to more-epitaxial processes. Gaps in the transparency windows of a-$SiO_2$ near 2.8 μm, a-SiN near ~3.0 μm, and a-Si near ~5.0 μm arise from O–H, N–H, and Si–H absorptions, respectively. (**b**) Magnified view of the highlighted region, using linear scales for both $n$ and transparency wavelength.

## Infrared transparent materials for thick windows

Unlike thin films, optical windows require low absorption over millimeter-scale propagation lengths. Accordingly, to consider a material transparent for window applications in **Fig. 2**, we applied a criterion of $\alpha < 1$ $cm^{-1}$ (compared to $\alpha < 10$ $cm^{-1}$ for thin films), corresponding to an internal transmittance of > 99% at 500 μm thickness.

Under this definition, materials that can serve both as thin films and bulk windows have narrower transparency windows compared to what is shown in **Fig. 1** (e.g., look at the curves for a-$SiO_2$ or c-Si). In **Fig. 2**, we also include several materials not considered for **Fig. 1**, including diamond [68], [69], halides (KBr [70], NaCl [71], and KRS-5 [72]), and complex oxides/nitrides (aluminum oxynitride (ALON) [73] and $MgAl_2O_4$ [74]); while these materials are challenging to deposit as thin films, there are well-established processes to synthesize them in bulk form, and they are widely used for IR windows.

### (i) Fluoride insulators (c-$MgF_2$, poly-$CaF_2$, poly-$BaF_2$)

Fluoride materials have nearly identical transparency windows from UV to IR wavelengths under both the thin-film and bulk window criteria. Both single crystals and polycrystals are widely used for bulk windows, enabled by Czochralski [16], Bridgman [18], or hot-pressing methods [75]. Note that c-$MgF_2$ crystallizes in the rutile (tetragonal) structure and is therefore birefringent ($\Delta n = n_e - n_o \sim 0.01$ at $\lambda = 2$ μm, where $n_e$ and $n_o$ are the extraordinary and ordinary refractive indices, respectively), limiting its use in polarization-sensitive applications. The anisotropy can be avoided using hot-pressed poly-$MgF_2$, where the random orientation of small crystalline domains results in isotropic optical properties, and this material is typically used for windows [76].

In contrast, $CaF_2$ and $BaF_2$ adopt the cubic fluorite structure and are therefore optically isotropic even in single-crystal form. The optical constants for $CaF_2$ and $BaF_2$ reported in **Fig. 2** were measured on polycrystalline samples, which are typically produced by hot pressing and meet optical-grade performance for window materials [75]. Grain boundaries can degrade optical performance, but the biggest effect of grain boundaries is scattering in the UV, without much of an impact on the mid IR due to the grain-boundary dimensions being far smaller than mid-IR wavelengths [77], [78].

Although $BaF_2$ has a broader transparency window than other fluorides, it is mechanically weaker than $CaF_2$ and poorly resistant to thermal and mechanical shock [76]. In addition, $BaF_2$ is more susceptible to hydrolysis at elevated temperatures in the presence of water vapor (onset at ~500 °C [79] vs. ~600 °C for $CaF_2$ [80]), and requires careful handling due to barium toxicity concerns. $CaF_2$ is therefore the more widely used IR window material due to its well-characterized environmental durability and lower production cost [81].

**(ii) Oxide insulators (a-$SiO_2$, c-$SiO_2$ [quartz], c-$Al_2O_3$, poly-$Y_2O_3$, c-$TiO_2$)**

Oxide-based windows provide excellent mechanical hardness and chemical stability, but their IR transparency is limited, especially under the bulk-window criteria. For example, a-$SiO_2$ is transparent to $\lambda < 5$ µm considering our thin-film coating criterion ($\alpha < 10$ cm$^{-1}$), but its IR cutoff shortens to $\lambda < 3.6$ µm using the window criterion ($\alpha < 1$ cm$^{-1}$). Typically, the transmittance of oxides cuts off between $\lambda = 3.5$–8 µm, much shorter wavelengths than that of fluoride, halide, or diamond windows, as shown in **Fig. 2**. These oxide windows are typically fabricated by flame fusion or vapor-phase synthesis for a-$SiO_2$ [82], hydrothermal growth for c-$SiO_2$ [83], Kyropoulos growth for c-$Al_2O_3$(sapphire) [84], Verneuil growth for c-$TiO_2$ (rutile) [85], and hot-pressing and sintering of fine ceramic powders for poly-$Y_2O_3$ [86].

a-$SiO_2$ [29] is one of the most cost-effective window materials, and processes such as flame fusion or vapor-phase synthesis allow independent control of metallic impurity and hydroxide (–OH) content [87]. For example, in UV-grade a-$SiO_2$, metallic impurities are minimized to improve UV transmission, while hydroxide absorption can be observed at $\lambda = 1.4$ µm, 2.2 µm, and 2.7 µm, where the 2.7 µm vibrational mode acts as the dominant absorption feature [87], and results in a small gap ($\lambda = 2.6$–2.9 µm) in the transparency window shown in **Fig. 2**. These hydroxide vibrational absorption features can be minimized by reducing the –OH content, and a-$SiO_2$ with –OH content below 1 ppm is considered to be "IR grade" [87]. Note that even without reducing hydroxide levels, –OH absorption in typical a-$SiO_2$ is too weak to appear as transparency gaps when using the thin-film criterion (**Fig. 1**) as opposed to the window criterion (**Fig. 2**).

Crystalline quartz (c-$SiO_2$), grown hydrothermally, exhibits birefringence ($\Delta n < 0.01$ at $\lambda = 2$ µm) due to its trigonal structure, making it suitable for broadband waveplates operating from the UV to the near-IR [88], [89]. Additionally, c-$SiO_2$ is sometimes chosen over a-$SiO_2$ for window applications that require higher surface hardness, higher thermal conductivity, or lower hydroxide content [88]; in such cases, z-cut c-$SiO_2$ is typically used for window applications to avoid polarization-dependent effects. However, slow hydrothermal growth rates and higher cost limit the scalability of c-$SiO_2$, and a-$SiO_2$ therefore remains the more widely used window material for both UV and near-IR applications [18].

c-$Al_2O_3$ (sapphire) [90] has a slightly broader transparency window, up to $\lambda < 5$ µm, and outstanding mechanical durability and thermal shock resistance, making it a standard short-wave and mid-wave IR window material. Although sapphire is optically anisotropic ($\Delta n \sim 0.01$ at $\lambda = 2$ µm), commercial windows are typically *c*-axis-plane cut so that normal-incident light propagates along the optical axis (experiencing $n_o$ for both polarizations), minimizing birefringence effects in most applications [91]. However, birefringence effects can still arise for off-axis or focused beams and must be accounted for in precision optical systems.

Poly-$Y_2O_3$ (yttria) [92] offers a wider transparency window, extending to $\lambda \sim 8$–9 µm, and high thermal stability (melting temperature ~2400 °C), making it a promising high-temperature IR window material [86]. However, its lower mechanical strength and thermal-shock resistance compared to sapphire and/or complex oxide/nitrides such as ALON and $MgAl_2O_4$ (spinel) limit its applications [93].

Finally, c-$TiO_2$ (rutile) [94] exhibits the highest *n* among oxides but its transparency is limited to $\lambda < 5$ µm. Furthermore, its tetragonal crystal structure results in birefringence ($\Delta n \sim 0.25$ at $\lambda = 1$ µm) much larger than that of sapphire. As a result, *c*-cut (001) orientation ($n_o$) is typically chosen for window uses, whereas *a*-cut (100) orientation, providing this large birefringence, is preferred for polarizing beam splitters and wave plates. We note that c-$TiO_2$ also exists in the anatase phase, but anatase irreversibly transforms to rutile above 600 °C, well below the temperature required for bulk single-crystal growth, and therefore rutile phase is the common phase for bulk windows [95].

**(iii) Complex oxide/nitride insulators (poly-ALON, poly-$MgAl_2O_4$)**

Complex oxides/nitrides such as ALON ($Al_{23}O_{27}N_5$) [73] and spinel ($MgAl_2O_4$) [74] have emerged as advanced alternatives to sapphire for IR windows, offering a transparency range of $\lambda = 0.4$–5 µm. Despite their multi-component compositions, both adopt cubic crystal structures, which render them optically isotropic with $n \sim 1.75$ in their polycrystalline form.

ALON and spinel are both fabricated via powder processing techniques such as cold isostatic pressing (performed at room temperature) [96], or hot isostatic pressing (performed at ~1800 °C) [97], [98] combined with sintering, which allows complex conformal geometrics such as domes and curved windows to be shaped directly during powder compaction [96]. In contrast, single-crystal sapphire must first be grown as a large ingot and then shaped into the desired form, a costly process that limits the achievable geometries.

**(iv) Halide insulators (c-KBr, c-NaCl, c-KRS5)**

Halide-based window materials include alkali halides such as potassium bromide (KBr) [70], sodium chloride (NaCl) [71], and thallium bromo-iodide ($TlBr_{0.4}I_{0.6}$, commonly known as KRS5) [72], all of which exhibit exceptional transparency across the IR regime. c-NaCl is transparent up to $\lambda < 20$ µm, c-KBr $\lambda < 28$ µm, and c-KRS5 extends further to $\lambda \sim 40$ µm (although our plots in **Fig. 2** are limited to the 0.2–25 µm range). Notably, both KBr and NaCl are also transparent into the UV region. NaCl and KBr have modest refractive index of 1.5–1.55, while KRS5 has a higher index at 2.3–2.4.

c-KBr and c-NaCl are typically grown by the Kyropoulos technique with relatively low cost [18]. However, both materials are mechanically soft, highly water-soluble, and strongly hygroscopic, requiring operation under low-humidity conditions. In contrast, c-KRS5, which is typically grown by the Bridgman–Stockbarger technique [18], is non-hygroscopic and has low water solubility, and has higher mechanical strength than the other halides. Because thallium compounds are highly toxic [99], handling of KRS5 must be performed with extreme care.

**(v) Chalcogenide semiconductors (poly-ZnS, poly-ZnSe, c-CdTe)**

ZnS and ZnSe are the most widely used chalcogenide window materials. Poly-ZnS is transparent from 0.4 to ~7 µm [46], [47] and poly-ZnSe from 0.6 to 18 µm [25], with a slightly higher *n* (~ 2.2 for ZnS and ~2.4 for ZnSe), as shown in **Fig. 2**. Both can be grown in polycrystalline form by CVD on graphite substrates, enabling large areas and lower cost than their single-crystalline forms while yielding good homogeneity and optical quality [49]. Notably, ZnS offers higher mechanical hardness and environmental durability than ZnSe [49].

c-CdTe (**Fig. 2**) [48] also exhibits broad IR transparency comparable to that of ZnSe. However, its practical adoption as a window material has been limited primarily by both the difficulty of growing defect-free single crystals by vertical Bridgman [100] even though it is the most preferred method, and its mechanical fragility. Furthermore, cadmium (Cd) is toxic, and substantial research has focused on replacing it with tin (Sn) [50]. Se is also toxic, though less so than Cd, and similarly requires handling precautions.

**(vi) Group III–V semiconductors (c-GaP, c-GaAs)**

GaP [52] and GaAs [53] are transparent in the mid-IR, with transparency windows of $\lambda < 6$ µm and $\lambda < 17$ µm, respectively. Notably, GaP is one of the highest-*n* materials transparent in the visible ($n \sim 3$–3.3, with a transparency window for $\lambda > 0.6$ µm). Both GaP and GaAs possess high thermal conductivity and strong thermal shock resistance, making them well suited for high-power IR laser systems (e.g., $CO_2$ lasers), where mechanical robustness and heat dissipation are critical [101]. Bulk GaP and GaAs can be readily grown as large single crystals using Czochralski or Vertical Gradient Freeze (VGF) methods [101], making them suitable for IR windows and lenses.

**(vii) Group IV semiconductors (c-Si, c-Ge)**

c-Si and c-Ge are the 'classic' mid-IR materials, widely used for mid-IR windows due to their low dispersion and mechanical robustness. They are typically produced as large, high-purity crystals via Czochralski [16] or float-zone growth [60].

In bulk form, the transmission of c-Si is limited to below $\lambda < 10$ µm due to phonon resonances at longer wavelengths, with some additional transparency windows between $\lambda = 14$–14.2 µm and $\lambda = 21$–24 µm [60]. c-Ge is transparent for $\lambda < 16$ µm [66]. As shown in **Fig. 2**; these transparency windows are significantly narrower than using the thin-film criterion in **Fig. 1**.

Since both c-Si and c-Ge are narrow-bandgap semiconductors, their performance can degrade at elevated temperatures due to thermally induced free-carrier absorption [102], [103]. This effect is far more pronounced in Ge due to its smaller bandgap (0.6 eV for c-Ge vs. 1.1 eV for c-Si). For example, K. Desnijder et al. demonstrated that c-Ge transparency in the mid-IR is noticeably reduced even slightly above room temperature and becomes nearly opaque above ~170 °C, which can limit its use in high-temperature optical systems [103].

**(viii) Diamond**

c-Diamond exhibits an exceptionally broad transparency window, extending from ~0.24 to ~2.2 µm and again from ~5.7 µm and to beyond 25 µm [68], [69]; the gap in between is caused by C–C phonon absorption. Diamond has a moderately high $n \sim 2.4$, and is one of the most durable infrared materials available due to its highest hardness and thermal conductivity among bulk materials and excellent resistance to thermal shock.

Optical-grade diamond windows are typically fabricated by CVD [54], which enables the growth of large, high-purity single crystalline or polycrystalline diamond. However, the extreme hardness of diamond may cause challenges in polishing after growth [104], often requiring advanced methods such as UV laser polishing [105] or microwave plasma-assisted polishing [106] to achieve smooth surfaces.

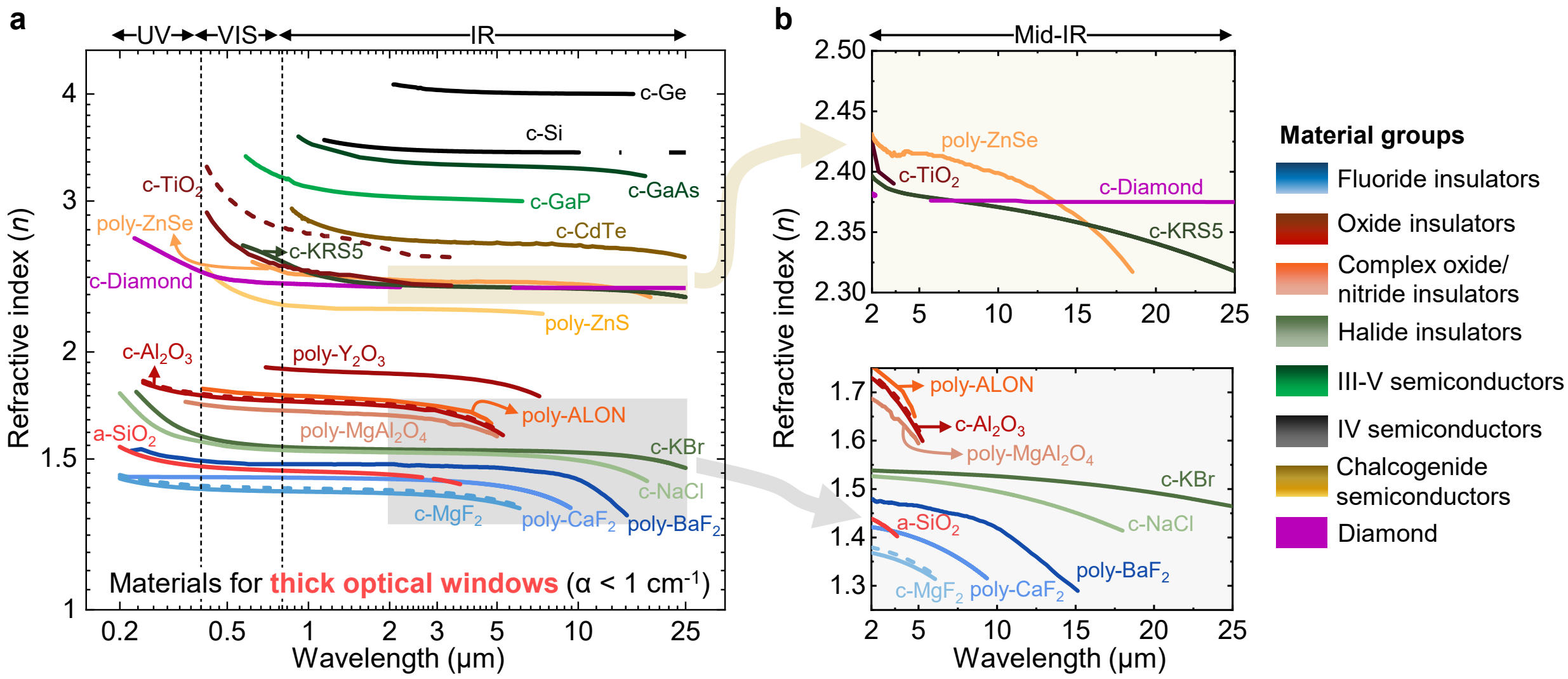

**Fig. 2. Infrared-transparent materials for thick infrared windows.** (**a**) Real part of the refractive index ($n$) for various candidate materials, where the wavelength regions satisfying an absorption coefficient of $\alpha < 1$ cm$^{-1}$ are defined as their transparency window and plotted over the 0.2–25 μm spectral range. Unlike in Fig. 1, complex oxide/nitride insulators, halide insulators, and diamond are included as distinct material groups due to their availability as window materials. Dashed lines correspond to the extraordinary refractive index ($n_e$) for birefringent crystals (c-$MgF_2$, c-$Al_2O_3$, and c-$TiO_2$), whereas solid lines correspond to the ordinary index ($n_o$) or isotropic index. Note that diamond and a-$SiO_2$ have gaps in their transparency window. (**b**) Magnified views of the highlighted regions in (a) where multiple materials overlap; In (b), both axes are plotted on a linear scale.

## Discussion and limitations

The optical bandgap ($E_g$) of a material determines its short-wavelength absorption edge, and is inversely correlated to the refractive index $n$, which quantifies the electronic polarizability of the medium. For example, fluoride materials ($MgF_2$, $CaF_2$, or $BaF_2$) have large bandgaps ($E_g > 10$ eV) arising from their strongly ionic bonding, resulting in low electronic polarizability and thus low refractive indices ($n < 1.5$). In contrast, materials with small bandgaps ($E_g < 1.2$ eV), such as Ge and Si, exhibit covalent bonding, where high electronic polarizability leads to higher refractive indices ($n > 3.0$).

In 1950, T. S. Moss proposed a simple empirical relation, showing the connection between $E_g$ and $n$: $n^4 E_g = C$, where $C$ is 95 eV [107]. In **Fig. 3**, we plot experimental $E_g$ *vs*. $n^*$ for IR-transparent materials, where $n^*$ is the real part of the complex index observed at the wavelength of lowest dispersion within each material's transparency range. Our dataset generally follows the overall trend of the Moss relation, though the experimental relationship is not perfectly captured by the relation. There have been several efforts to refine or extend the Moss relation [107], [108], [109], though it is clear from examining **Fig. 3** that no single $E_g$ vs $n$ curve can fit all existing materials.

Although extensive optical measurements have been performed on most common window and thin-film optical materials (**Figs. 1–3**), there remain plenty of experimental unknowns even in the most common materials. For example, all of the optical properties summarized in this review were obtained at room temperature. In practice, many mid-IR applications require high-temperature conditions, where the temperature dependence of the optical constants—both the refractive index ($dn/dT$) and the extinction coefficient ($d\kappa/dT$)—becomes a design consideration. For example, in high-power laser systems, thermally induced refractive index changes cause beam distortion [110], transmission loss [102], [103], or unexpected thermal runaway [111]. In thermal-emitter applications, such as thermophotovoltaics [112] and high-temperature infrared sources [113], the temperature-dependent optical constants affect the emissivity spectrum. Some temperature-dependent datasets are available, for example a-$SiO_2$ for $\lambda$ = 5–25 μm from room temperature to ~870 K [87], c-$CaF_2$, c-$BaF_2$, and c-$MgF_2$ for $\lambda$ = 0.2–1.7 µm from ~300 to ~640 K [114], poly-ZnS for $\lambda$ = 0.5–14 µm from 93 to 1000 K, poly-ZnSe for $\lambda$ = 0.55–18 µm from 93 to 618 K, and poly-ZnTe for $\lambda$ = 0.5–30 µm at room temperature [115], intrinsic c-Si for $\lambda$ = 14–140 µm for 300 to 450 K [116], but additional studies are needed to cover many IR-transparent materials over broad wavelength and temperature ranges, particularly for the applications discussed above.

This review paper is accompanied by datasets [19] including the real and imaginary refractive index spectra for all materials discussed, along with scripts to generate plots such as those in **Figs. 1** and **2**. In this main text, we plot the transparency windows only for $\alpha < 10$ $cm^{-1}$ for thin film coatings in **Fig. 1** and $\alpha < 1$ $cm^{-1}$ for bulk windows in **Fig. 2**, but users with different requirements can generate custom transparency maps for any threshold value of $\alpha$, using the provided scripts [19].

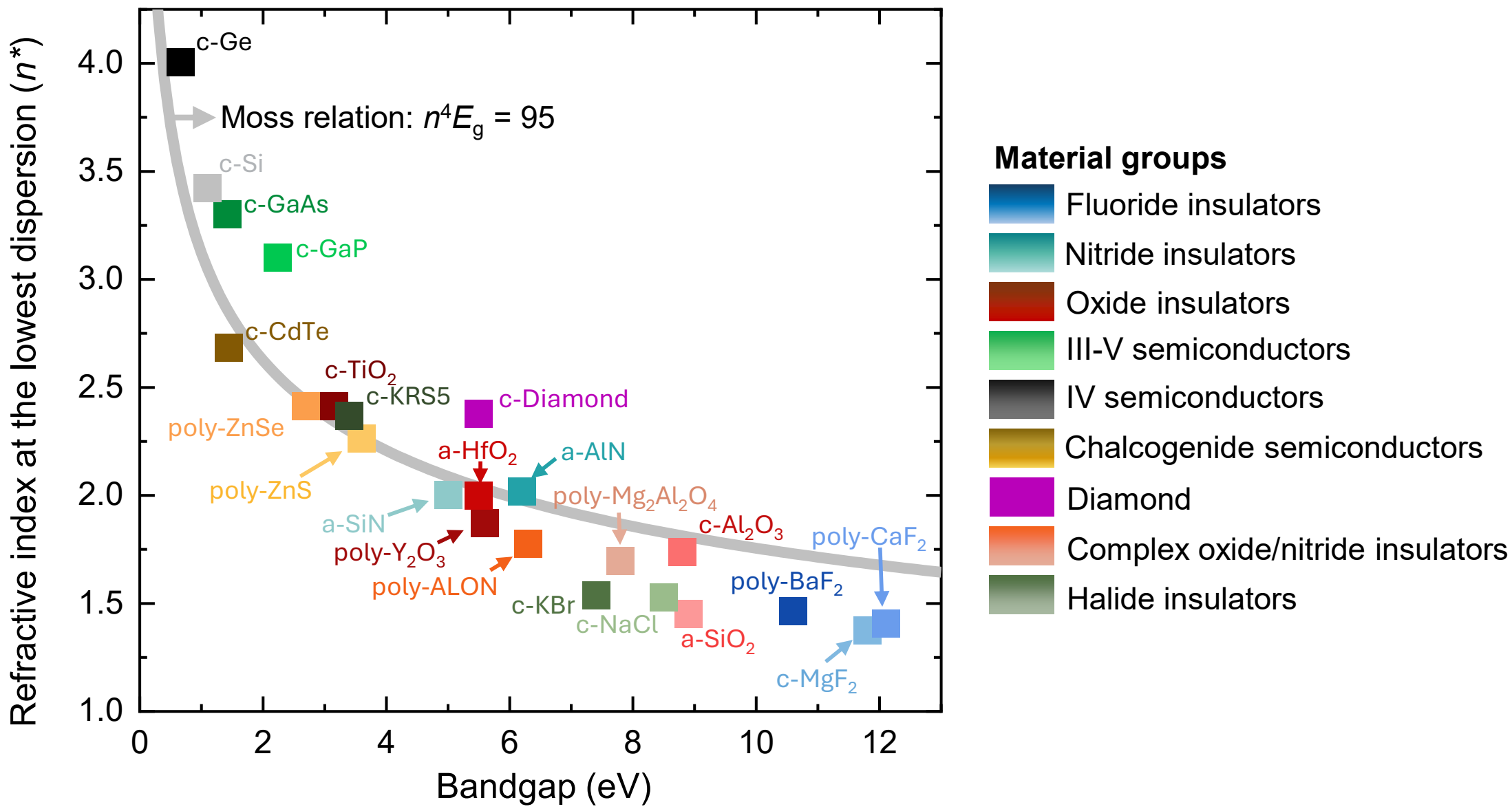

**Fig. 3. Correlation between the refractive index ($n^*$) and the optical bandgap ($E_g$).** For each selected material, $n^*$ was taken at the wavelength point of lowest dispersion. The Moss relation ($n^4E_g$ = 95 eV) is also plotted, as a reference curve.

## Acknowledgements

We acknowledge support from the UW-Madison Grainger Institute of Engineering. T. Kumar acknowledges support from the NASA FINESST grant (80NSSC25K0314) and David G. Walsh Travel Award from the McPherson Eye Research Institute.

## Conflict of Interest

The authors declare no conflict of interest.

## Data Availability Statement

The optical constants ($n$, $\kappa$) for all materials discussed in this work are publicly available via Zenodo at http://doi.org/10.5281/zenodo.20187375, along with the analysis scripts used to generate the figures [19].

## References

[1] M. Hou, Y. Chen, J. Li, and F. Yi, “Single 5-centimeter- aperture metalens enabled intelligent lightweight mid-infrared thermographic camera,” *Sci. Adv.*, vol. 10, no. 27, pp. 1–12, 2024, doi: 10.1126/sciadv.ado4847.

[2] S. Rosas, W. Adi, A. Beisenova, S. K. Biswas, F. Kuruoglu, H. Mei, M. A. Kats, D. A. Czaplewski, Y. S. Kivshar, and F. Yesilkoy, “Enhanced biochemical sensing with high-Q transmission resonances in free-standing membrane metasurfaces,” *Optica*, vol. 12, no. 2, pp. 178–189, 2025, doi: 10.1364/optica.549393.

[3] B. Fang, S. C. Bodepudi, F. Tian, X. Liu, D. Chang, S. Du, J. Lv, J. Zhong, H. Zhu, H. Hu, Y. Xu, Z. Xu, W. Gao, and C. Gao, “Bidirectional mid-infrared communications between two identical macroscopic graphene fibres,” *Nat. Commun.*, vol. 11, 6368, 2020, doi: 10.1038/s41467-020-20033-2.

[4] Y. Yao, A. J. Hoffman, and C. F. Gmachl, “Mid-infrared quantum cascade lasers,” *Nat. Photonics,* vol. 6, no. 7, pp. 432–439, 2012, doi: 10.1038/nphoton.2012.143.

[5] A. John-Herpin, A. Tittl, L. Kühner, F. Richter, S. H. Huang, G. Shvets, S.-H. Oh, and H. Altug, “Metasurface-Enhanced Infrared Spectroscopy: An Abundance of Materials and Functionalities,” *Adv. Mater.*, vol. 35, no. 34, 2023, doi: 10.1002/adma.202110163.

[6] L. Zhang, A. M. Agarwal, L. C. Kimerling, and J. Michel, “Nonlinear Group IV photonics based on silicon and germanium: From near-infrared to mid-infrared,” *Nanophotonics*, vol. 3, no. 4–5, pp. 247–268, 2014, doi: 10.1515/nanoph-2013-0020.

[7] H. Shim, Z. Kuang, and O. Miller, “Optical materials for maximal nanophotonic response [Invited],” *Opt. Mater. Express*, vol. 10, no. 7, pp. 1561–1585, 2020, doi: 10.1364/ome.396419.

[8] P. Klocek (Ed.), Handbook of infrared optical materials (1st ed.), CRC Press, 1991, doi: 10.1201/9781315213996.

[9] “The Correct Material for Infrared (IR) Applications.” [Online]. Available: https://www.edmundoptics.com/knowledge-center/application-notes/optics/the-correct-material-for-infrared-applications/?srsltid=AfmBOopmSOzptoVtDAjPteBRjVfk6sPoSl7tEouxtdsi32KHsmmwIzXm

[10] “What optical materials work best in the IR (infrared)?” [Online]. Available: https://escooptics.com/blogs/news/what-optical-materials-work-best-in-the-ir-infrared

[11] S. Yue, Y. Liu, R. Wang, Y. Hou, H. Shi, Y. Feng, Z. Wen, and Z. Zhang, “All-silicon polarization-independent broadband achromatic metalens designed for the mid-wave and long-wave infrared,” *Opt. Express*, vol. 31, no. 26, p. 44340, 2023, doi: 10.1364/oe.506471.

[12] H. I. Lin, J. Geldmeier, E. Baleine, F. Yang, S. An, Y. Pan, C. Rivero-Baleine, T. Gu, J. Hu, “Wide-Field-of-View, Large-Area Long-Wave Infrared Silicon Metalenses,” *ACS Photonics*, vol. 11, no. 5, pp. 1943–1949, 2024, doi: 10.1021/acsphotonics.4c00013.

[13] J. S. Yu, S. Jung, J.-W. Cho, G.-T. Park, M. Kats, S.-K. Kim, E. Lee, “Ultrathin Ge-$YF_3$ antireflective coating with 0.5% reflectivity on high-index substrate for long-wavelength infrared cameras,” *Nanophotonics*, vol. 13, no. 21, pp. 4067–4078, 2024, doi: 10.1515/nanoph-2024-0360.

[14] F. Kuruoglu, S. Rosas, J.-W. Cho, D. A. Czaplewski, Y. Kivshar, M. Kats and F. Yesilkoy, "Dynamically Tunable Membrane Metasurfaces for Infrared Spectroscopy," pp. 1–15, 2025, doi:10.48550/arXiv.2506.10115.

[15] L. Norder, S. Yin, M. H. J. de Jong, F. Stallone, H. Aydogmus, P. M. Sberna, M. A. Bessa, and R. A. Norte, "Pentagonal photonic crystal mirrors: scalable lightsails with enhanced acceleration via neural topology optimization," *Nat. Commun.*, vol. 16, 2753, pp. 1–11, 2025, doi: 10.1038/s41467-025-57749-y.

[16] J. Friedrich, W. von Ammon, and G. Müller, Czochralski Growth of Silicon Crystals, Second Edi., vol. 2. Elsevier B.V., 2015. doi: 10.1016/B978-0-444-63303-3.00002-X.

[17] M. Bruel, "Silicon on insulator material technology," Electron. Lett., vol. 31, no. 14, pp. 1201–1202, 1995, doi: 10.1049/el:19950805.

[18] W. J. Tropf, M. E. Thomas, and P. Klocek, "Infrared optical materials," Proc. SPIE, vol. 10286, pp. 102860A, 1992, doi: 10.1117/12.245197.

[19] J.-W. Cho, T. Kumar, H. Mei, M. A. Kats (2026). IR transparency window. Zenodo. https://doi.org/10.5281/zenodo.20187375

[20] M. A. Kats, R. Blanchard, P. Genevet, and F. Capasso, "Nanometre optical coatings based on strong interference effects in highly absorbing media," *Nat. Mater.*, vol. 12, no. 1, pp. 20–24, 2013, doi: 10.1038/nmat3443.

[21] R. A. Outlaw and P. Hopson Jr., "Free-standing thin film Ge single crystals grown by plasma-enhanced chemical vapor deposition," *J. Appl. Phys.*, vol. 55, pp. 1461–1463, 1984, doi: 10.1063/1.333401.

[22] K. A. McComber, X. Duan, J. Liu, J. Michel, and L. C. Kimerling, "Single-crystal germanium growth on amorphous silicon," *Adv. Funct. Mater.*, vol. 22, no. 5, pp. 1049–1057, 2012, doi: 10.1002/adfm.201102015.

[23] D. Franta, D. Nečas, A. Giglia, P. Franta, and I. Ohlídal, "Universal dispersion model for characterization of optical thin films over wide spectral range: Application to magnesium fluoride," *Appl. Surf. Sci.*, vol. 421, pp. 424–429, 2017, doi: 10.1016/j.apsusc.2016.09.149.

[24] D. F. Bezuidenhout, "Calcium Fluoride ($CaF_2$)," Handb. Opt. Constants Solids, vol. II, pp. 815–835, 1997, doi: 10.1016/B978-012544415-6.50082-0.

[25] M. R. Querry, "Optical constants of minerals and other materials from the millimeter to the ultraviolet, Contractor Report CRDEC-CR-88009, " pp. 61–70, 1987.

[26] J. C. Phillips and J. A. Van Vechten, "Dielectric classification of crystal structures, ionization potentials, and band structures," *Phys. Rev. Lett.*, vol. 22, no. 14, pp. 705–708, 1969, doi: 10.1103/PhysRevLett.22.705.

[27] H. J. Lee, D. Wang, T. H. Kim, D.-H. Jung, T.-H. Kil, K.-S. Lee, H.-J. Choi, S.-H. Baek, E. Yoon, W. J. Choi, and J. M. Baik, "Wide-temperature (up to 100 °C) operation of thermostable vanadium oxide based microbolometers with Ti/$MgF_2$ infrared absorbing layer for long wavelength infrared (LWIR) detection," *Appl. Surf. Sci.*, vol. 547, p. 149142, 2021, doi: 10.1016/j.apsusc.2021.149142.

[28] T. Yoshida, K. Nishimoto, K. Sekine, and K. Etoh, “Fluoride antireflection coatings for deep ultraviolet optics deposited by ion-beam sputtering,” *Appl. Opt.*, vol. 45, no. 7, pp. 1375–1379, 2006, doi: 10.1364/AO.45.001375.

[29] D. Franta, D. Nečas, I. Ohlídal, and A. Giglia, “Optical characterization of $SiO_2$ thin films using universal dispersion model over wide spectral range,” *Proc. SPIE Opt. Micro- Nanometrology VI*, vol. 9890, p. 989014, 2016, doi: 10.1117/12.2227580.

[30] D. Franta, D. Nečas, I. Ohlídal, and A. Giglia, “Dispersion model for optical thin films applicable in wide spectral range,” *Proc. SPIE Opt. Syst. Des. 2015 Opt. Fabr. Testing, Metrol. V*, vol. 9628, p. 96281U, 2015, doi: 10.1117/12.2190104.

[31] D. Franta, J. Vohánka, and M. Čermák, “Universal dispersion model for characterization of optical thin films over a wide spectral range: application to hafnia,” *Appl. Opt.*, vol. 54, no. 31, pp. 9108–9119, 2015, doi: 10.1007/978-3-319-75325-6_3.

[32] M. Lax and E. Burstein, “Infrared lattice absorption in ionic and homopolar crystals,” *Phys. Rev.*, vol. 97, no. 1, pp. 39–52, 1955, doi: 10.1103/PhysRev.97.39.

[33] R. Ellerbrock, M. Stein, and J. Schaller, “Comparing amorphous silica, short-range-ordered silicates and silicic acid species by FTIR,” *Sci. Rep.*, vol. 12, no. 1, pp. 1–8, 2022, doi: 10.1038/s41598-022-15882-4.

[34] J. Xu, S. Li, W. Zhang, S. Yan, C. Liu, X. Yuan, X. Ye, and H. Li, “The impact of deposition and annealing temperature on the growth properties and surface passivation of silicon dioxide films obtained by atomic layer deposition,” *Appl. Surf. Sci.*, vol. 544, no. December 2020, p. 148889, 2021, doi: 10.1016/j.apsusc.2020.148889.

[35] K. Hofmann, G. W. Rubloff, and R. A. McCorkle, “Defect formation in thermal $SiO_2$ by high-temperature annealing,” *Appl. Phys. Lett.*, vol. 49, no. 22, pp. 1525–1527, 1986, doi: 10.1063/1.97322.

[36] J.-W. Cho, Y.-J. Lee, J.-H. Kim, R. Hu, E. Lee, and S.-K. Kim, “Directional Radiative Cooling via Exceptional Epsilon-Based Microcavities,” *ACS Nano*, vol. 17, no. 11, pp. 10442–10451, 2023, doi: 10.1021/acsnano.3c01184.

[37] K. Luke, Y. Okawachi, M. R. E. Lamont, A. L. Gaeta, M. Lipson, “Broadband mid-infrared frequency comb generation in a $Si_3N_4$ microresonator,” *Opt. Lett.*, vol. 40, no. 21, pp. 4823–4826, 2015, doi: 10.1364/OL.40.004823.

[38] L. Y. Beliaev, E. Shkondin, A. V. Lavrinenko, and O. Takayama, “Optical, structural and composition properties of silicon nitride films deposited by reactive radio-frequency sputtering, low pressure and plasma-enhanced chemical vapor deposition,” *Thin Solid Films*, vol. 763, pp. 139568, 2022, doi: 10.1016/j.tsf.2022.139568.

[39] T. Al Moussi, C. O’Dalaigh, P. Raynaud, J. Esvan, P. Lambkin, R. Lakshmanan, B. Chen, and S. Diaham, “Structural, optical and electrical properties of Si-rich and N-rich PECVD silicon nitride films,” *Sci. Rep.*, vol. 15, no. 1, pp. 1–14, 2025, doi: 10.1038/s41598-025-14296-2.

[40] A. Frigg, A. Boes, G. Ren, I. Abdo, D.-Y. Choi, S. Gees, and A. Mitchell, “Low loss CMOS-compatible silicon nitride photonics utilizing reactive sputtered thin films,” *Opt. Express*, vol. 27, no. 26, pp. 37795–37805, 2019, doi: 10.1364/oe.380758.

[41] S. Jafari, J. Hirsch, D. Lausch, M. John, N. Bernhard, and S. Meyer, “Composition limited hydrogen effusion rate of a-SiNx:H passivation stack,” *AIP Conf. Proc.*, vol. 2147, 2019, doi: 10.1063/1.5123853.

[42] T. Kumar, D. Feng, S. Yin, M. Mah, P. Lin, M. A. Fortman, G. R. Jaffe, C. Wan, H. Mei, Y. Xiao, R. Synowicki, R. J. Warzoha, V. W. Brar, J. J. Talghader, and M. A. Kats, “Self-Referencing Photothermal Common-Path Interferometry to Measure Absorption of $Si_3N_4$ Membranes for Laser-Light Sails,” *ACS Photonics*, vol. 12, no. 11, pp. 6381–6387, 2025, doi: 10.1021/acsphotonics.5c01886.

[43] L. Y. Beliaev, E. Shkondin, A. V. Lavrinenko, and O. Takayama, “Thickness-dependent optical properties of aluminum nitride films for mid-infrared wavelengths*,” J. Vac. Sci. Technol. A Vacuum, Surfaces, Film.*, vol. 39, no. 4, 2021, doi: 10.1116/6.0000884.

[44] H. Fu, X. Huang, H. Chen, Z. Lu, and Y. Zhao, “Fabrication and Characterization of Ultra-wide Bandgap AlN-Based Schottky Diodes on Sapphire by MOCVD*,” IEEE J. Electron Devices Soc.*, vol. 5, no. 6, pp. 518–524, 2017, doi: 10.1109/JEDS.2017.2751554.

[45] H. Miyake, C. H. Lin, K. Tokoro, and K. Hiramatsu, “Preparation of high-quality AlN on sapphire by high-temperature face-to-face annealing,” *J. Cryst. Growth*, vol. 456, pp. 155–159, 2016, doi: 10.1016/j.jcrysgro.2016.08.028.

[46] J.-W. Cho, E.-J. Lee, and S.-K. Kim, “Full-Color Solar-Heat-Resistant Films Based on Nanometer Optical Coatings,” *Nano Lett.*, 2021, doi: 10.1021/acs.nanolett.1c04043.

[47] H. Mei, J.-W. Cho, J.-S. Yu, H. Chen, S. Singh, B. Zhao, J. Ravichandran, S.-K. Kim, M. A. Kats, “Anti-reflection coatings for highly anisotropic materials in the mid infrared,” arXiv:2601.00558, doi: 10.48550/arXiv.2601.00558.

[48] E. D. Palik, “Cadmium Telluride (CdTe),” Handb. Opt. Constants Solids, vol. I, pp. 409–427, 1997, doi: 10.1016/B978-012544415-6.50017-0.

[49] D. W. Blodgett, M. E. Thomas, D. V. Hahn, and S. G. Kaplan, “Longwave infrared absorption and scatter properties of ZnS and ZnSe,” *Proc. SPIE Wind. Dome Technol. VIII*, vol. 5078, p. 137, 2003, doi: 10.1117/12.487874.

[50] S. J. Wakeham and G. J. Hawkins, “Investigation of cadmium alternatives in thin-film coatings,” *Proc. SPIE Adv. Thin-Film Coatings Opt. Appl. III*, vol. 6286, p. 62860C, 2006, doi: 10.1117/12.680349.

[51] N. Zakay, A. Schlesinger, U. Argaman, L. Nguyen, N. Maman, B. Koren, M. Ozeri, G. Makov, Y. Golan, and D. Azulay, “Electrical and Optical Properties of γ-SnSe: A New Ultra-narrow Band Gap Material,” *ACS Appl. Mater. Interfaces*, vol. 15, no. 12, pp. 15668–15675, 2023, doi: 10.1021/acsami.2c22134.

[52] S. Adachi, "Gallium Phosphide (GaP)," in Optical Constants of Crystalline and Amorphous Semiconductors: Numerical Data and Graphical Information. Boston, MA: Kluwer Academic, Springer Science & Business Media, pp. 198–212, 1999.

[53] E. D. Palik, "Gallium Arsenide (GaAs)," Handb. Opt. Constants Solids, vol. I, pp. 429–443, 1997, doi: 10.1016/b978-012544415-6.50018-2.

[54] S. Nakamura, "III-V nitride based light-emitting devices," *Solid State Commun.*, vol. 102, no. 2–3, pp. 237–243, 1997, doi: 10.1016/S0038-1098(96)00722-3.

[55] A. W. Bett, F. Dimroth, G. Stollwerck, and O. V. Sulima, "III-V compounds for solar cell applications," *Appl. Phys. A Mater. Sci. Process.*, vol. 69, no. 2, pp. 119–129, 1999, doi: 10.1007/s003390050983.

[56] H. Emmer, C. T. Chen, R. Saive, D. Friedrich, Y. Horie, A. Arbabi, A. Faraon, and H. A. Atwaterm, "Fabrication of Single Crystal Gallium Phosphide Thin Films on Glass," *Sci. Rep.*, vol. 7, no. 1, pp. 2–7, 2017, doi: 10.1038/s41598-017-05012-w.

[57] H. U. Chae, B. Shrewsbury, R. Ahsan, M. L. Povinelli, and R. Kapadia, "GaAs Mid-IR Electrically Tunable Metasurfaces," *Nano Lett.*, vol. 24, no. 8, pp. 2581–2588, 2024, doi: 10.1021/acs.nanolett.3c04687.

[58] D. A. Mota, G. H. Chandra, J. Ventura, A. Guedes, and J. P. De Cruz, "Influence of Process Parameters on the RF Sputtered GaP Thin Films," *J. Mater. Sci. Technol.*, vol. 29, no. 9, pp. 821–829, 2013, doi: 10.1016/j.jmst.2013.06.005.

[59] M. Imaizumi, M. Adachi, Y. Fujii, Y. Hayashi, T. Soga, T. Jimbo, M. Umeno, "Low-temperature growth of GaAs polycrystalline films on glass substrates for space solar cell application," *J. Cryst. Growth*, vol. 221, pp. 688–692, 2000, doi: doi:10.1016/S0022-0248(00)00801-0.

[60] D. Franta, A. Dubroka, C. Wang, A. Giglia, J. Vohánka, P. Franta, and I. Ohlídal, "Temperature-dependent dispersion model of float zone crystalline silicon," *Appl. Surf. Sci.*, vol. 421, pp. 405–419, 2017, doi: 10.1016/j.apsusc.2017.02.021.

[61] R. J. Collins and H. Y. Fan, "Infrared lattice absorption bands in germanium, silicon, and diamond," *Phys. Rev.*, vol. 93, no. 4, pp. 674–678, 1954, doi: 10.1103/PhysRev.93.674.

[62] M. Fox, *Optical Properties of Solids*, 2nd ed. Oxford: Oxford University Press, 2010.

[63] D. Franta, D. Nečas, L. Zajíčková, I. Ohlídal, and J. Stuchlík, "Advanced modeling for optical characterization of amorphous hydrogenated silicon films," *Thin Solid Films*, vol. 541, pp. 12–16, 2013, doi: 10.1016/j.tsf.2013.04.129.

[64] D. Franta, D. Nečas, L. Zajíčková, I. Ohlídal, J. Stuchlík, and D. Chvostová, "Application of sum rule to the dispersion model of hydrogenated amorphous silicon," *Thin Solid Films*, vol. 539, pp. 233–244, 2013, doi: 10.1016/j.tsf.2013.04.012.

[65] A 100-nm-thick a-Ge was deposited on a c-Si substrate by e-beam evaporation, and its optical constants were measured by IR-VASE (infrared variable-angle spectroscopic ellipsometry, J.A. Woollam) and fitted using WVASE software.

[66] R. F. Potter, "Germanium (Ge)," Handb. Opt. Constants Solids, vol. I, pp. 465–478, 1997, doi: 10.1016/B978-012544415-6.50020-0.

[67] M. Kim, M. Zahedian, W. Wu, C. Fang, Z. Yu, R. A. Wambold, R. Vidrio, Y. Tong, S. Yin, D. A. Czaplewski, J. T. Choy, and M. A. Kats, “Broadband Light Extraction from Near-Surface NV Centers Using Crystalline-Silicon Antennas,” *Nano Lett.*, vol. 25, no. 12, pp. 4659–4666, 2025, doi: 10.1021/acs.nanolett.4c04299.

[68] D. F. Edwards and E. Ochoa, “Infrared refractive index of diamond,” *J. Opt. Soc. Am.*, vol. 71, no. 5, pp. 607–608, 1981, doi: 10.1364/josa.71.000607.

[69] A. M. Bennett, B. J. Wickham, H. K. Dhilon, Y. Chen, S. Webster, G. Turri, and M. Bass, “Development of high-purity optical grade single-crystal CVD diamond for intracavity cooling,” *Proc. SPIE Solid State Lasers XXIII Technol. Devices*, vol. 8959, p. 89590R, 2014, doi: 10.1117/12.2037811.

[70] E. D. Palik, “Potassium Bromide (KBr),” Handb. Opt. Constants Solids, vol. II, pp. 989–1004, 1997, doi: 10.1016/B978-012544415-6.50090-X.

[71] J. E. Eldridge and E. D. Palik, “Sodium chloride (NaCl),” Handb. Opt. Constants Solids, vol. I, pp. 775–793, 1997, doi: 10.1016/B978-012544415-6.50041-8.

[72] W. J. Tropf, “Cubic Thallium (I) Halides,” Handb. Opt. Constants Solids, vol. III, pp. 923–961, 1997, doi: 10.1016/B978-012544415-6.50140-0.

[73] W. J. Tropf and M. E. Thomas, “Aluminum Oxynitride (ALON) Spinel,” Handb. Opt. Constants Solids, vol. II, pp. 777–787, 1997, doi: 10.1016/B978-012544415-6.50079-0.

[74] W. J. Tropf and M. E. Thomas, “Magnesium Aluminum Spinel ($MgAl_2O_4$),” Handb. Opt. Constants Solids, vol. II, pp. 883–894a, 1997, doi: 10.1016/B978-012544415-6.50086-8.

[75] E. Carnall, S. E. Hatch, and W. F. Parsons, “Optical Studies on Hot-Pressed Polycrystalline $CaF_2$ With Clean Grain Boundaries,” *Role Grain Boundaries Surfaces Ceram.*, pp. 165–173, 1966, doi: 10.1007/978-1-4899-6311-6_11.

[76] D. C. Harris, *Materials for Infrared Windows and Domes: Properties and Performance*. Bellingham, WA, USA: SPIE Press, 1999, doi: 10.1117/3.349896.

[77] “Photonchina.” [Online]. Available: https://www.photonchinaa.com/product/caf2/

[78] “FarfieldCrystal”, [Online]. Available: https://fairfieldcrystal.com/the-strengths-of-magnesium-fluoride-in-uv-optics/

[79] “Barium Fluoride Windows.” [Online]. Available: https://www.thorlabs.com/barium-fluoride-windows/?gad_source=1&gad_campaignid=21369512767&gbraid=0AAAAAD_iN3gAKfjqFLocp77gvRhLQnPbE&gclid=CjwKCAjwwJzPBhBREiwAJfHRnZrM5wzmbe_H9uyTSL8FUsQMsd040bAgbGiBeOpnAFW7uJt1X71fHRoCvwoQAvD_BwE&tabName=Overview

[80] “Calcium Fluoride Windows.” [Online]. Available: https://www.thorlabs.com/calcium-fluoride-windows?tabName=Overview

[81] D. Hahn, “Calcium Fluoride and Barium Fluoride Crystals in Optics: Multispectral optical materials for a wide spectrum of applications,” *Opt. Photonik*, vol. 9, no. 4, pp. 45–48, 2014, doi: 10.1002/opph.201400066.

[82] R. Brockner, “Properties and structure of vitreous silica,” *J. Non-Cryst. Solids*, vol. 5, no. pp. 123–175 January, 1970, doi: 10.1016/0022-3093(70)90190-0.

[83] R. J. Baughman, “Quartz crystal growth,” *J. Cryst. Growth*, vol. 112, no. 4, pp. 753–757, 1991, doi: 10.1016/0022-0248(91)90132-O.

[84] I. Kryvonosov, E. Dolzhenkova, P. Konevskiy, L. Lytvynov, and A. Voloshyn, “Growing of large-sized high-quality sapphire by the Kyropoulos method,” *Funct. Mater.*, vol. 32, no. 1, pp. 5–12, 2025, doi: 10.15407/fm32.01.5.

[85] X. Liu, H. Ma, L. Wang, Y. Hu, and X. Sun, “Growth of $TiO_2$ single crystals by the Verneuil method at different gas flow ratio,” *J. Cryst. Growth*, vol. 623, no. June, p. 127403, 2023, doi: 10.1016/j.jcrysgro.2023.127403.

[86] C. T. Mathew, S. Solomon, J. Koshy, and J. K. Thomas, “Infrared transmittance of hybrid microwave sintered yttria,” *Ceram. Int.*, vol. 41, no. 8, pp. 10070–10078, 2015, doi: 10.1016/j.ceramint.2015.04.100.

[87] S. Yin, J.-W. Cho, D. Feng, H. Mei, T. Kumar, C. Wan, Y. Jin, M. Kim, and M. A. Kats, “Preventing overfitting in infrared ellipsometry using temperature dependence: fused silica as a case study,” *Opt. Mater. Express*, vol. 15, no. 8, pp. 1939–1948, 2025, doi: 10.1364/ome.564694.

[88] “Crystal Quartz $SiO_2$.” [Online]. Available: https://www.crystran.com/optical-materials/crystal-quartz-sio2/

[89] “Synthetic Crystal Quartz.” [Online]. Available: https://www.tydexoptics.com/materials1/for_transmission_optics/crystal_quartz/

[90] F. Gervais, “Aluminum Oxide ($Al_2O_3$),” Handb. Opt. Constants Solids, vol. II, pp. 761–775, 1997, doi: 10.1016/B978-012544415-6.50078-9.

[91] “Sapphire Windows.” [Online]. Available: https://www.knightoptical.com/stock/windows-and-diffusers/ir-windows/sapphire-windows?utm_term=&hsa_acc=2814001386&hsa_cam=23200485728&hsa_grp=&hsa_ad=&hsa_src=x&hsa_tgt=&hsa_kw=&hsa_mt=&hsa_net=adwords&hsa_ver=3&gad_source=1&gad_campaignid=23204752513&gbraid=0AAAACL0LOR0o4kwMAUtTbtaRJs3qB0k8

[92] W. J. Tropf and M. E. Thomas, “Yttrium Oxide ($Y_2O_3$),” Handb. Opt. Constants Solids, vol. II, pp. 1079–1096, 1997, doi: 10.1016/b978-012544415-6.50096-0.

[93] P. Hogan, T. Stefanik, C. Willingham, R. Gentilman, R. Integrated, and D. Systems, “Transparent Yttria for IR Windows and Domes – Past and Present,” *10th DoD Electromagn. Wind. Symp.*, 2004.

[94] M. W. Ribarsky, “Titanium Dioxide ($TiO_2$),” Handb. Opt. Constants Solids, vol. I, pp. 795–804, 1997, doi: 10.1016/B978-012544415-6.50042-X.

[95] D. A. H. Hanaor and C. C. Sorrell, “Review of the anatase to rutile phase transformation,” *J. Mater. Sci.*, vol. 46, no. 4, pp. 855–874, 2011, doi: 10.1007/s10853-010-5113-0.

[96] L. M. Goldman, M. Smith, M. Ramisetty, S. Jha, and S. Sastri, "Conformal ALON and spinel windows," *Proc. SPIE Wind. Dome Technol. XVI*, vol. 10985, no. 109850G, p. 16, 2019, doi: 10.1117/12.2518902.

[97] G. Gilde, P. Patel, P. Patterson, D. Blodgett, D. Duncan, and D. Hahn, "Evaluation of hot pressing and hot isostastic pressing parameters on the optical properties of spinel," *J. Am. Ceram. Soc.*, vol. 88, no. 10, pp. 2747–2751, 2005, doi: 10.1111/j.1551-2916.2005.00527.x.

[98] N. Jiang, Q. Liu, T. Xie, P. Ma, H. Kou, Y. Pan, J. Li, "Fabrication of highly transparent AlON ceramics by hot isostatic pressing post-treatment," *J. Eur. Ceram. Soc.*, vol. 37, no. 13, pp. 4213–4216, 2017, doi: 10.1016/j.jeurceramsoc.2017.04.028.

[99] "ISP Optics Thallium Bromoiodide (KRS-5) Windows." [Online]. Available : https://www.edmundoptics.com/f/thallium-bromoiodide-krs-5-windows/39725/?srsltid=AfmBOophxforWh3VPwLyA98p8kjo20kxVyfdB1uHzBzJDLNk10DoXmVn

[100] P. Rudolph and M. Mühlberg, "Basic problems of vertical Bridgman growth of CdTe," *Mater. Sci. Eng. B*, vol. 16, no. 1–3, pp. 8–16, 1993, doi: 10.1016/0921-5107(93)90005-8.

[101] J. Václavík and D. Vápenka, "Gallium Phosphide as a material for visible and infrared optics," *EPJ Web of Conferences*, vol. 48, pp. 1–4, 2013, doi: 10.1051/epjconf/20134800028.

[102] H. A. Weakliem and D. Redfield, "Temperature dependence of the optical properties of silicon," *J. Appl. Phys.*, vol. 50, no. 3, pp. 1491–1493, 1979, doi: 10.1063/1.326135.

[103] K. Desnijder, G. D. Lewis, and M. Vandewal, "Optical degradation of germanium with elevated temperature : Consequences for infrared imaging systems," *Opt. Laser Technol.*, vol. 192, no. PC, p. 113648, 2025, doi: 10.1016/j.optlastec.2025.113648.

[104] T. Schuelke and T. A. Grotjohn, "Diamond polishing," *Diam. Relat. Mater.*, vol. 32, pp. 17–26, 2013, doi: 10.1016/j.diamond.2012.11.007.

[105] H. Liu, L. Xie, W. Lin, and M. Hong, "Optical Quality Laser Polishing of CVD Diamond by UV Pulsed Laser Irradiation," vol. 2100537, pp. 1–8, 2021, doi: 10.1002/adom.202100537.

[106] X. Li, Y. Xiao, Y. Wang, Q. He, and Y. Zhang, "Microwave plasma-assisted polishing of polycrystalline diamond," *Diam. Relat. Mater.*, vol. 152, no. 1088, p. 111907, 2025, doi: 10.1016/j.diamond.2024.111907.

[107] T. S. Moss, "Relations between the Refractive Index and Energy Gap of Semiconductors," *Phys. Status Solidi*, vol. 131, no. 2, pp. 415–427, 1985, doi: 10.1002/pssb.2221310202.

[108] R. Ravichandran, A. X. Wang, and J. F. Wager, "Solid state dielectric screening versus band gap trends and implications," *Opt. Mater.*, vol. 60, pp. 181–187, 2016, doi: 10.1016/j.optmat.2016.07.027.

[109] H. M. Gomaa, I. S. Yahia, and H. Y. Zahran, "Correlation between the static refractive index and the optical bandgap: Review and new empirical approach," *Phys. B Condens. Matter*, vol. 620, p. 413246, 2021, doi: 10.1016/j.physb.2021.413246.

[110] C. A. Klein, "Optical distortion coefficients of high-power laser windows," *Opt. Eng.*, vol. 29, no. 4, p. 343, 1990, doi: 10.1117/12.55600.

[111] G. R. Holdman, G. R. Jaffe, D. Feng, M. S. Jang, M. A. Kats, and V. W. Brar, "Thermal Runaway of Silicon-Based Laser Sails," *Adv. Opt. Mater.*, vol. 10, no. 19, pp. 1–7, 2022, doi: 10.1002/adom.202102835.

[112] R. Sakakibara, V. Stelmakh, W. R. Chan, M. Ghebrebrhan, J. D. Joannopoulos, M. Soljacic, and I. Čelanović, "Practical emitters for thermophotovoltaics: a review," *J. Photonics Energy*, vol. 9, no. 03, p. 1, 2019, doi: 10.1117/1.jpe.9.032713.

[113] O. Ilic, P. Bermel, G. Chen, J. D. Joannopoulos, I. Celanovic, and M. Soljačić, "Tailoring high-temperature radiation and the resurrection of the incandescent source," *Nat. Nanotechnol.*, vol. 11, no. 4, pp. 320–324, 2016, doi: 10.1038/nnano.2015.309.

[114] Q. Zheng, X. Wang, and D. Thompson, "Temperature-dependent optical properties of monocrystalline $CaF_2$, $BaF_2$, and $MgF_2$," *Opt. Mater. Express*, vol. 13, no. 8, p. 2380, 2023, doi: 10.1364/ome.496246.

[115] A. W. Elshaari, I. E. Zadeh, K. D. Jöns, and V. Zwiller, "Thermo-Optic Characterization of Silicon Nitride Resonators for Cryogenic Photonic Circuits," *IEEE Photonics J.*, vol. 8, no. 3, 2016, doi: 10.1109/JPHOT.2016.2561622.

[116] H. H. Li, "Refractive Index of ZnS, ZnSe, and ZnTe and Its Wavelength and Temperature Derivatives," *J. Phys. Chem. Ref. Data*, vol. 13, no. 1, pp. 103–150, 1984, doi: 10.1063/1.555705.